# Hydrogen storage in rippled graphene: perspectives from multi-scale simulations

Vito Dario Camiola[1,2], Riccardo Farchioni[1,2,3], Tommaso Cavallucci[2,3], Antonio Rossi[2,4], Vittorio Pellegrini[1,2,4], Valentina Tozzini[1,2]*

[1]Istituto Nanoscienze – Cnr, Piazza San Silvestro 12, 56127 Pisa Italy
[2]NEST – Scuola Normale Superiore, Piazza San Silvestro 12, 56127 Pisa Italy
[3]Physics Department Enrico Fermi, University of Pisa, Largo Bruno Pontecorvo Pisa Italy
[4]Graphene Labs, IIT Istituto Italiano di Technologia, via Morego, 30, 16163 Genova Italy

* **Correspondence:** Valentina Tozzini, valentina.tozzini@nano.cnr.it

Exploring new perspectives for green technologies is one of the challenges of the third millennium, in which the need for non-polluting and renewable powering has become primary. In this context, the use of hydrogen as a fuel is promising, since the energy released in its oxidation (~285 kJ/mole) is three times that released, on average, by hydrocarbons, and the combustion product is water (Ramage, 1983). Being hydrogen a vector of chemical energy, efficient conservation and non-dispersive transportation are the main goals. Three issues must be considered to this respect: (i) storage capacity (ii) storage stability (iii) kinetics of loading/release. Commercial technologies are currently based on cryo-compression or liquefaction of $H_2$ in tanks. These ensure quite a high gravimetric density (GD, point (i)), namely 8-13% in weight of stored hydrogen, and a relatively low cost (Züttel 2003). However concerning points (ii) and (iii), these technologies pose problems of safety, mainly due to explosive flammability of hydrogen, and consequent unpractical conditions for transportation and use (Mori et al 2009). Therefore, research efforts are directed towards solid-state based storage systems (energy.gov, Bonaccorso et al 2015).

Interactions of hydrogen with materials are classified as physisorption, occurring with $H_2$ by means of van der Waals (vdW) forces, or chemisorption, i.e. chemical binding of H leading to the formation of hydrides (Mori et al 2009), requiring dissociative(associative) chemi(de)sorption of $H_2$. Intermediate nature interactions, sometimes called "phenisorption", can also occur between hydrogen electrons and the electrons of external orbital of metals. Indeed, stable and robust (light) metal hydrides (Harder et al 2011, Sakintuna et al 2007) are currently considered an alternative to tanks. Their main drawback is the high chemi(de)sorption barrier, implying slow operational kinetics, which becomes acceptable only at very high temperature. Physisorption, conversely, generally results in barrierless and weak binding. It was considered as a storage mechanism in layered (Zhirko 2007) or porous (Sastre 2010) materials, and shown to be effective at low temperatures and/or high pressure. Therefore, it generally seems that if storage stability (ii) is improved then the loading/release kinetics (iii) is worsened.

Graphene shows good potential to be an efficient hydrogen storage medium (Tozzini V et al, 2013): carbon is among the lightest elements forming layered and porous structures, and graphene is probably the material with the largest surface to mass ratio. These two conditions are in principle optimal to produce high GD (point i). In addition, the chemical versatility of carbon allows it to interact with hydrogen both by physisorption (in sp2 hybridization) and chemisorption (Goler et al 2013) (in sp3 hybridization). ("Phenisorption" is also obtained by in graphene by functionalization with metals (Mashoff et al 2013)).

On the other hand, concerning points (ii) and (iii), pure graphene does not perform dramatically better than other materials. $H_2$ easily physisorbs onto graphene layers or within multilayers, but it was theoretically shown (Patchkovskiim et al 2005) that large GD (6-8%) are reached within multi-layered graphene at cryogenic temperatures, while the room temperature value is at best ~2-3%. This was confirmed by measurements (Klechikov et al submitted), which also indicate that graphene does not perform better than other carbon based bulk materials, such as nanoporous carbon or carbon nanotubes. In all cases, a key parameter determining GD is the specific surface to volume ratio. Theoretical works also show that stability can be improved (and GD optimized) at specific interlayer spacing (~7-8Å), due to a cooperative effect of vdW forces(Patchkovskiim et al 2005). A similar effect is responsible for the accumulation of physisorbed hydrogen within graphene troughs at low temperatures (~100K) observed in simulations(Tozzini et al 2011). On the other hand, hydrogen chemisorption on graphene produces graphane(Sofo et al 2007), its completely hydrated alkane counterpart, stable at room temperature (point (ii)) and with 8.2% GD (point (i)). Graphane, however, shares with other hydrides high chemi(de)sorption barrier (~1.5 eV/atom). As in other materials, physisorption has good kinetics (iii), and bad storage capacity (i) and stability (ii), while chemisorption has good (i) and (ii), and bad (iii).

Graphene, however, joins together extremely peculiar properties: a unique combination of strength and flexibility, and its bidimensionality. Therefore it can be buckled on different scales down to nanometer, (Wang ZF, 2011, Fasolino A, 2007), either statically, i.e. forming stable ripples as an effect of external constraints (i.e. compression or interaction with a substrate (Goler et al 2013)), or dynamically, sustaining traveling ripples, i.e. coherent transverse out-of-plane acoustic modes, also called ZA or flexural phonons(Lindsay et al 2010, Xu et al 2013). We explored the possibility of exploiting these properties in the context of hydrogen storage.

We first evaluated the dependence of chemisorbed H stability on the local curvature by means of a Density Functional Theory (DFT) based study(Tozzini et al 2011). This revealed a linear dependency of the binding energy on the local curvature, leading to an over-stabilization of the adsorbate of up to 1-2 eV on crests and corresponding destabilization within troughs (Fig1, A), which is also modified by the presence of already chemisorbed hydrogen (Rossi et al, submitted). Theoretical evaluations were confirmed by Scanning Tunneling Microscopy performed on naturally corrugated graphene grown on SiC(Goler et al 2013), which demonstrated the presence of hydrogen prevalently on the convexities. These observations lead to the idea that an inversion of curvature (from concavity to convexity) could detach chemisorbed hydrogen. As a matter of fact, this mechanism was demonstrated in a simulation in which the curvature inversion was realized dynamically by the passage of a ZA coherent phonon of ~2 nm wavelength and ~THz frequency (Fig 1, B). The simulation shows that hydrogen undergoes associative desorption at room temperature upon curvature inversion, which can be called a "mechanical catalysis": the energy barrier is overcome by means of the energy provided by the traveling phonon.

Once detached, $H_2$ finds itself in contact with a graphene sheet with traveling ripples. If sheets are at appropriate distances and if ZA phonons are excited with appropriate relative phases in subsequent sheets, the traveling ripples enclose traveling nano-cavities (Fig 1, C). We showed by means of classical Molecular Dynamics (MD) simulations using empirical Force Fields (FF) (Camiola, submitted) that these nano-cavities can include and transport $H_2$ at velocities near to the phase velocity of the phonon for almost micrometric distances before the phonon damps. This effect could be used to transport and pump hydrogen through multilayers, improving the GD of physisorbed hydrogen at room temperature and the kinetics of loading and release.

The existence of ripples and flexural phonons, therefore, offers specific strategies – unique to graphene – to overcome issues (i), (ii) and (iii). The proof of principle of the effectiveness of these strategies was given by means of a multi-scale approach, combining DFT and empirical FF based MD simulations, in order to couple high accuracy in the representation of interactions to the large time and size scales needed to represent the process. However, the realization of these processes relies on the practical possibility of finely manipulating the local curvature and generating and sustaining coherent flexural phonons. In fact, our current efforts are along that route: we are exploring several possibilities for static and dynamic curvature manipulation, including external electric fields(Cavallucci 2014), functionalization with optically active molecular pillars (Burress et al, 2010), electro-mechanical pulling, coupling to piezo-electric substrates (Camiola et al, submitted). In this phase a multi-scale simulation approach will be even more necessary: in the route towards the practical realization of possible devices, one must analyze also macroscopic effects of thermodynamic nature, in addition to the physical phenomena at the sub-nano scale and the dynamical behavior at the nano-to-micro scale. Therefore, we intend develop a continuum representation, where graphene is modeled as a membrane with mechano-elastic properties (Zang et al (2013)) to the DFT and to the MD with empirical FF. Consistency between the representations will be obtained by appropriate bottom-up parameterization and top-down transfer of macroscopic information, and will ensure a complete and accurate representation of the system properties and behavior.

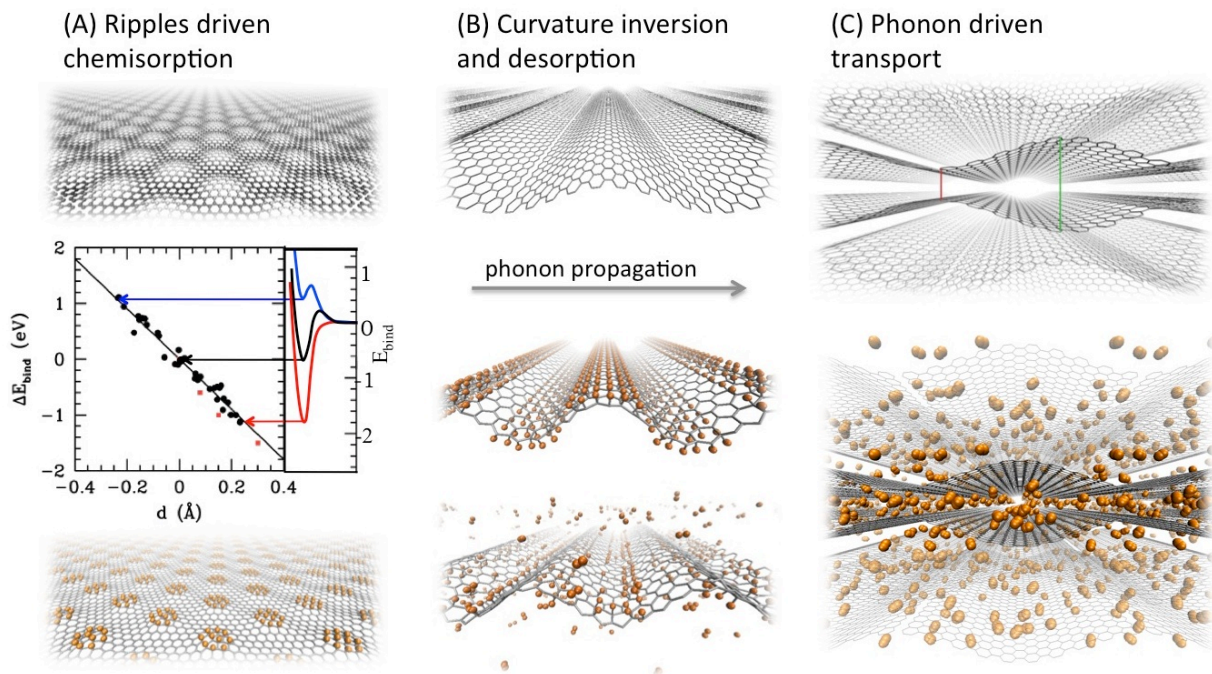

**Figure 1.** Hydrogen interaction with rippled and buckled graphene. (A) from top to bottom: graphene with buckling mimicking that of graphene on SiC; variation of stability *vs* curvature, measured by means of the out of plane displacement of a C site with respect to the neighbors, d (blue and negative value of *d* = concave, red and positive value of *d* = convex); hydrogen (shown in orange) binds on concavities. (B) From top to bottom: coherent flexural phonon of nano-sized wavelength; hydrogen is initially attached on the crest of the ripples, but after half a period troughs replace crests, hydrogen destabilizes and desorbs in molecular form. (C) From top to bottom: flexural phonons excited in counterphase in subsequent multilayers generate traveling cavities; hydrogen is transported by phonons within those cavities.

## Acknowledgement
We gratefully acknowledge financial support by the EU, 7th FP, Graphene Flagship (contract no. NECT-ICT-604391), the CINECA award "ISCRA C" IsC10_HBG, 2013 and PRACE "Tier0" award Pra07_1544, and IIT Compunet Platform for computational resources.